\documentclass[aps, pre, twocolumn, superscriptaddress, nofootinbib, amsmath, amssymb]{revtex4-1}

\usepackage{amsmath}
\usepackage{bm}
\usepackage{comment}
\usepackage{dcolumn}
\usepackage{graphicx}

\newcommand{\tr}{\ensuremath{\mathrm{Tr}}}

\begin{document}

\title{Topological effects in continuum 2d $U(N)$ gauge theories}

\author{Claudio Bonati}
\email{claudio.bonati@df.unipi.it}

\author{Paolo Rossi}
\email{paolo.rossi@unipi.it}

\affiliation{Dipartimento di Fisica, Universit\`a di Pisa and INFN, Sezione di
Pisa, Largo Pontecorvo 3, 56127 Pisa, Italy}

\date{\today}

\begin{abstract}
We study the $\theta$ dependence of the continuum limit of 2d $U(N)$ gauge
theories defined on compact manifolds, with special emphasis on spherical
($g=0$) and toroidal ($g=1$) topologies. We find that the coupling between
$U(1)$ and $SU(N)$ degrees of freedom survives the continuum limit, leading to
observable deviations of the continuum topological susceptibility from the $U(1)$
behavior, especially for $g=0$, in which case deviations remain even in the large
$N$ limit.
\end{abstract}

\maketitle

\section{Introduction}

It is well known that two dimensional gauge theories are analytically much more
tractable than their four dimensional counterparts, and in some cases some
exact nonperturbative expressions can even be obtained. Quite surprisingly,
however, the $\theta$ dependence of these two dimensional models appears not to
have been investigated until very recently. 

In the paper \cite{Bonati:2019ylr} we filled this gap, by presenting analytic
results for various aspects of the $\theta$ dependence of two dimensional
$U(N)$ gauge theories. By generalizing the argument presented in
\cite{Rusakov:1990rs} (see also \cite{Witten:1991we, Kiskis:2014lwa} for
different approaches), the partition function at $\theta\neq 0$ of the lattice
$U(N)$ gauge theory (with Wilson action) was written as sum over the
representations of $U(N)$ of some character's related coefficients. This
expression, although exact, is of little practical use due to its complexity,
so some limit cases were also investigated: the continuum limit at finite area
$A$, the thermodynamic limit $A\to\infty$ and the large $N$ limit in the
thermodynamic case (i.e. at $A=\infty$).

The two main outcomes of this analysis were the following: the first one is
that, in the thermodynamic limit, the large $N$ behaviour of the topological
susceptibility is different in the weak and strong coupling phases identified by the
Gross-Witten-Wadia transition \cite{Gross:1980he, Wadia:1980cp}. Indeed for 't
Hooft coupling $\beta<1/2$ the susceptibility diverges in the large $N$ limit,
while for $\beta>1/2$ its value is related to the expectation value of the
determinant of the link variables, as computed in \cite{Rossi:1982vw}.  The
second noteworthy result is a rather unexpected feature of the continuum
limit: against naive expectations based on continuum intuition, $SU(N)$ and
$U(1)$ degrees of freedom do not decouple from each other even in the continuum
limit as far as $A<\infty$. 

In this paper we will elaborate more on the second point, by rewriting the
continuum partition function in a form that makes manifest the interaction of
the $SU(N)$ degrees of freedom with the instanton sectors of the $U(1)$ theory.
We will also discuss the large $N$ limit at finite area and, in the case of
spherical topology ($g=0$), we will present numerical evidence that the
topological susceptibility behaves as an order parameter for the
Douglas-Kazakov transition \cite{Douglas:1993iia}.

\section{The partition function in the continuum limit}\label{sec:continuum}

In \cite{Bonati:2019ylr} it was shown that, starting from the Wilson action and the
discretization 
\begin{equation}
q(U_p)=-\frac{i}{2\pi} \tr \ln U_p\ ,
\end{equation}
of the topological charge density ($U_p$ is the parallel transporter around a
plaquette), the $\theta$ dependent, finite area partition function of the
continuum $U(N)$ model can be written in the form
\begin{equation}\label{eq:continuumpart}
Z_{\theta}^{(g)} (N, X) = \sum_{\{l_j\}} d_{\{l_j\}}^{2-2g} 
e^{-{X \over 2N}\big[C_{\{l_j\}}+ {\theta \over \pi }\sum_j l_j + {N \over 4 \pi^2}\theta^2 \big] }\ . 
\end{equation}
Here the $N$ integer numbers $\{l_j\}$ (with $l_1\ge l_2\ge \cdots\ge l_N$) label
the representations of $U(N)$ (see e.g. \cite{Drouffe:1983fv}), $g$ is the
genus of the manifold on which the theory is defined and $X= A/2\beta$ is a
dimensionless variable, depending on the area $A$ of the manifold and on the 't
Hooft coupling $\beta$. $d_{\{l_j\}}$ and $C_{\{l_j\}}$ denote the dimension and
the quadratic Casimir of the representation identified by $\{l_j\}$, whose
explicit expressions are
\begin{equation}
\begin{aligned}
C_{\{l_j\}}&=\sum_{j=1}^N l_j(l_j-2j+N+1) = \\
&= \sum_{j=1}^N \left(l_j-j+\frac{N+1}{2}\right)^2 -\frac{N(N^2-1)}{12}\ , \\
d_{\{l_j\}}&=\prod_{k>j}\left(1-\frac{l_k-l_j}{k-j}\right)\ .
\end{aligned}
\end{equation}

It is important to note that the expression in Eq.~\eqref{eq:continuumpart} is
consistent with the periodicity in $\theta$ of the partition function, with
period $2 \pi$. Indeed the exponents appearing in Eq.~\eqref{eq:continuumpart}
can be rewritten in the form
\begin{equation}
\begin{aligned}
& C_{\{l_j\}}+\frac{\theta}{\pi}\sum_j l_j + \frac{N}{4\pi^2}\theta^2 = \\
& =\sum_{j=1}^N \left(l_j +\frac{\theta}{2 \pi}-j+\frac{N+1}{2}\right)^2 -\frac{N(N^2-1)}{12}\ ;
\end{aligned}
\end{equation}
as a consequence $\theta\to\theta+2\pi$ is equivalent to $\{l_j\}\to\{l'_j\}$
where $l'_j=l_j+1$. Since $d_{\{l'_j\}} = d_{\{l_j\}}$ the $2\pi-$periodicity
of Eq.~\eqref{eq:continuumpart} immediately follows.

The particular case of the $U(1)$ gauge theory is obviously the simplest one:
in this case the partition function does not depend on the genus $g$ of the
manifold and it is simply given by
\begin{equation}\label{eq:ZU1}
Z_{\theta}(1, X) = \sum_{n} e^{-\frac{X}{2}\left(n+\frac{\theta}{2\pi}\right)^2}\ ,
\end{equation}
a result that can be readily obtained using more conventional methods (see e.g.
\cite{Cao:2013na}).

The topological susceptibility $\chi_t^{(g)}(N, \beta, A)$ can be
computed by using the general relation
\begin{equation}
\chi_t^{(g)}(N, \beta, A)=-\frac{1}{A}\frac{\partial^2}{\partial\theta^2} \log Z_{\theta}^{(g)} (N, X) 
\end{equation}
and, to make the notation more compact, it is convenient to define the
(normalized) weights
\begin{equation}\label{eq:cont_weights}
w_{\{l_j\}}^{(g)} \big(N, X \big) = d_{\{l_j\}}^{2-2g} e^{-{X \over 2 N} C_{\{l_j\}} } \Bigl[Z_0^{(g)}  \big(N, X \big) \Bigr]^{-1}.
\end{equation}
The finite volume continuum limit of the topological susceptibility (at
$\theta=0$) is then given by
\begin{equation}
\chi_t^{(g)}(N, \beta, A) = \frac{1}{8\pi^2 \beta}\Big[1 - X \sum_{\{l_j\}} w_{\{l_j\}}^{(g)} \Big(\sum_j \frac{l_j}{N} \Big)^2\Big]\ ,
\end{equation}
where the relation 
\begin{equation}
\sum_{\{l_j\}} \Big(w_{\{l_j\}}^{(g)} \sum_j l_j\Big)=0
\end{equation}
was used to simplify the result. This relation holds true since for each
representation $\{l_j\}$ the conjugate representation $\{l'_j\}$ (with
$l'_j=-l_{N+1-j}$) has the same weight of $\{l_j\}$ and $\sum_j l'_j=-\sum_j
l_j$.

In the infinite volume limit $X\to\infty$ it is easily seen that
$w_{\{l_j\}}^{(g)}\to \delta_{\{l_j\}, \{0\}}$ (where $\{0\}$ denotes the
trivial representation), and in this limit the topological susceptibility does
not depend on the genus $g$ and on the number of colors $N$, becoming
simply equal to
\begin{equation}
\chi_t(N, \beta, \infty) = \frac{1}{8 \pi^2 \beta}\ .
\end{equation}
Hence from now on, in order to simplify the notation, we shall express our
results for the topological susceptibility in terms of the
dimensionless ratio
\begin{equation}\label{eq:normsusc}
R^{(g)}(N, X)\ \equiv \frac{\chi_t^{(g)}(N, \beta, A)}{\chi_t(N, \beta, \infty)}\ .
\end{equation}
In some cases it will be useful to study also the derivative of $R^{(g)}(N, X)$
with respect to the area-related parameter $X$.  An explicit expression for
this quantity is
\begin{equation}\label{eq:dRdx}
\begin{aligned}
&{\partial R^{(g)}(N, X) \over \partial X} = \frac{X}{2N} \Bigg[ \sum_{\{l_j\}} w_{\{l_j\}}^{(g)} C_{\{l_j\}} \Big(\sum_j \frac{l_j}{N} \Big)^2 - \\
&-\Bigg(\sum_{\{l_j\}} w_{\{l_j\}}^{(g)} C_{\{l_j\}}\Bigg)\Bigg(\sum_{\{l_j\}} w_{\{l_j\}}^{(g)} \Big(\sum_j \frac{l_j}{N}\Big)^2 \Bigg) \Bigg] -\\
&-\sum_{\{l_j\}} w_{\{l_j\}}^{(g)} \Big(\sum_j \frac{l_j}{N} \Big)^2.\
\end{aligned}
\end{equation}

With the aim of clarifying the interaction between the $SU(N)$ and the $U(1)$
degrees of freedom, it is convenient to notice that representations of $U(N)$
can be unambiguously obtained from the representations of $SU(N)$ (see e.g.
\cite{Drouffe:1983fv}). In order to better exploit the symmetry between
representations and their conjugates we
change the summation index from $j$ to $i$, by setting
\begin{equation} 
i = j-{N+1 \over 2},
\end{equation}
where $j\in \{1,\ldots,N\}$ and the (integer or half-integer) numbers
$i$ runs from $-{1 \over 2}(N-1)$ to ${1 \over 2}(N-1)$.  Representations of
$SU(N)$ can be labelled by the (integer or half-integer) numbers $m_i = l_i-i$,
with the condition $m_{i} > m_{i+1}$ and an additional (conventional)
constraint fixing the value of one of the $m_i$ in order to avoid double
counting (we can for example fix $m_{\frac{N-1}{2}}=-\frac{N-1}{2}$, which is
equivalent to the condition $l_N=0$ used in \cite{Drouffe:1983fv}). The
representations of $U(N)$ will then be obtained from those of $SU(N)$ by the
substitutions $\{m_i\} \rightarrow \{m_i + n\}$, for all $n \in \mathbb{Z}$. 

To rewrite the partition function we observe that the relation between the
quadratic Casimir of $U(N)$ (denoted by $C_{\{l_j\}}$) and the corresponding
one of $SU(N)$ (denoted by $C_{\{m_i\}}$) is
\begin{equation}
C_{\{m_i\}} = C_{\{l_j\}}- \frac{1}{N} \Big( \sum_{j=1}^N l_j\Big)^2,
\end{equation}
and, since $\sum_i m_i = \sum_j l_j$, we have
\begin{equation}
\frac{1}{N} C_{\{m_i\}} = \frac{1}{N}\sum_i m_i^2 -\Big(\frac{1}{N} \sum_i m_i\Big)^2 -\frac{N^2-1}{12} \ .
\end{equation}
Moreover the relation between the dimensions of the representations is
\begin{equation}
d_{\{m_i\}} = \prod_{k>i} \left(\frac{m_i-m_k}{k-i}\right) = d_{\{l_j\}}\ .
\end{equation}
These observations allow to decompose the summation on $\{l_j\}$ in
Eq.~\eqref{eq:continuumpart} into a summation on $\{m_i\}$ and a summation on
$n$: it is easy to show that, by applying the above decomposition, the
partition function may be expressed as
\begin{equation}\label{eq:zetaNX}
\begin{aligned}
&\hspace{3cm} Z_\theta^{(g)}(N, X) = \\
&= \sum_{\{m_i\}} d^{2-2g}_{\{m_i\}} e^{-\frac{X}{2 N} C_{\{m_i\}}} \sum_n e^{-{X \over 2} (n+ {\theta \over 2\pi}+{1 \over N}\sum_i  m_i)^2} .
\end{aligned}
\end{equation}

It is now convenient to group the representations of $SU(N)$ according to the value taken
by $\sum_i m_i$. We then define the following $SU(N)-$ related functions
\begin{equation}
W^{(g)} (N,X,M) \equiv \sum_{\{m_i;M\}} d^{2-2g}_{\{m_i\}} e^{-\frac{X}{2 N} C_{\{m_i\}}},
\end{equation}
where the notation $\{m_i;M\}$ means that the sum is restricted to the
representations ${\{m_i\}}$ such that $\sum_i m_i = M$. The heat-kernel
partition function of $SU(N)$ is then given by
\begin{equation}
Z^{(g)}_{SU}(N, X) = \sum_M W^{(g)} (N,X,M)\ ,
\end{equation}
while the $U(N)$ partition function in Eq.~\eqref{eq:zetaNX} can be
rewritten, using the $U(1)$ partition function Eq.~\eqref{eq:ZU1}, as
\begin{equation}\label{eq:Zpartial}
Z_\theta^{(g)}(N, X) = \sum_M W^{(g)} (N,X,M)\, Z_{\theta+\mu} (1,X),
\end{equation}
where we introduced the notation
\begin{equation}
\mu \equiv 2\pi M/N \ .
\end{equation}
We can then exploit the Poisson formula to write  $Z_{\theta+\mu} (1,X)$ as
\begin{equation}
\sum_n e^{-\frac{X}{2} \big(n+ \frac{\theta +\mu}{2\pi}\big)^2} = \sqrt{\frac{2\pi}{X}} \sum_k e^{-\frac{2\pi^2 k^2}{X}+ik(\theta +\mu)},
\end{equation}
where $k$ labels the $k$-instanton configuration of the $U(1)$ vacuum (see e.g.
\cite{Cao:2013na}). It is now possible to exchange the order of summations in
Eq.~\eqref{eq:Zpartial} and obtain the representation
\begin{equation}\label{eq:zetaweak}
Z_\theta^{(g)}(N, X) = \sqrt{2\pi \over X}\sum_k e^ {-\frac{2 \pi^2 k^2}{X}+i k \theta}\, \tilde W^{(g)} (N,X,k),
\end{equation}
where
\begin{equation}\label{eq:sunpart}
\tilde W^{(g)} (N,X,k) = \sum_M e^{i k \mu}\,W^{(g)} (N,X,M) 
\end{equation}
is the Fourier transform of $W^{(g)} (N,X,M)$ and can be interpreted as the
partition function of the $SU(N)$ degrees of freedom in the $k$ instanton
$U(1)$ sector.

It is worth noticing that Eq.~\eqref{eq:zetaweak}, due to its simple dependence
on $\theta$, leads easily to an alternative formula for the evaluation of the
topological susceptibility, especially useful for the case in which $X$ is small, 
since only few $k$ values contribute significantly to the sum in this case.

The function $\tilde W^{(g)} (N,X,k)$ can be exactly computed in various
limits. When $X \rightarrow \infty$ the trivial representation dominates and
$W^{(g)}(N,X,M)\to \delta_{M,0}$; as a consequence $\tilde{W}^{(g)}(N,X,M)\to
1$ in this limit. When $g \rightarrow \infty$ representations of dimension 1
dominate the sums and again $W^{(g)}(N,X,M)\to \delta_{M,0}$ (due to the
constraint $m_{\frac{N-1}{2}}=-\frac{N-1}{2}$) and $\tilde{W}^{(g)}(N,X,M)\to
1$.  In the next section we will show that the same happens when $N \rightarrow
\infty$ with genus $g>1$. In all these limits the partition function reduces to
that of the $U(1)$ model, and as a consequence the same happens to the
topological susceptibility. We thus have for the ratio defined in
Eq.~\eqref{eq:normsusc}
\begin{equation}
R^{(g)}(N,X)\to R(X) \equiv 1 - X\frac{\sum_n n^2 e^{-\frac{X}{2} n^2} }{\sum_n e^{-\frac{X}{2} n^2}}\ ,
\end{equation}
and the universal function $R(X)$ satisfies the duality property
\begin {equation}
R(X) + R\left(\frac{4 \pi^2}{X}\right) = 1\ ,
\end{equation}
as can be seen by using the Poisson summation formula.

\begin{figure}[t] 
\centering 
\includegraphics[width=0.9\columnwidth, clip]{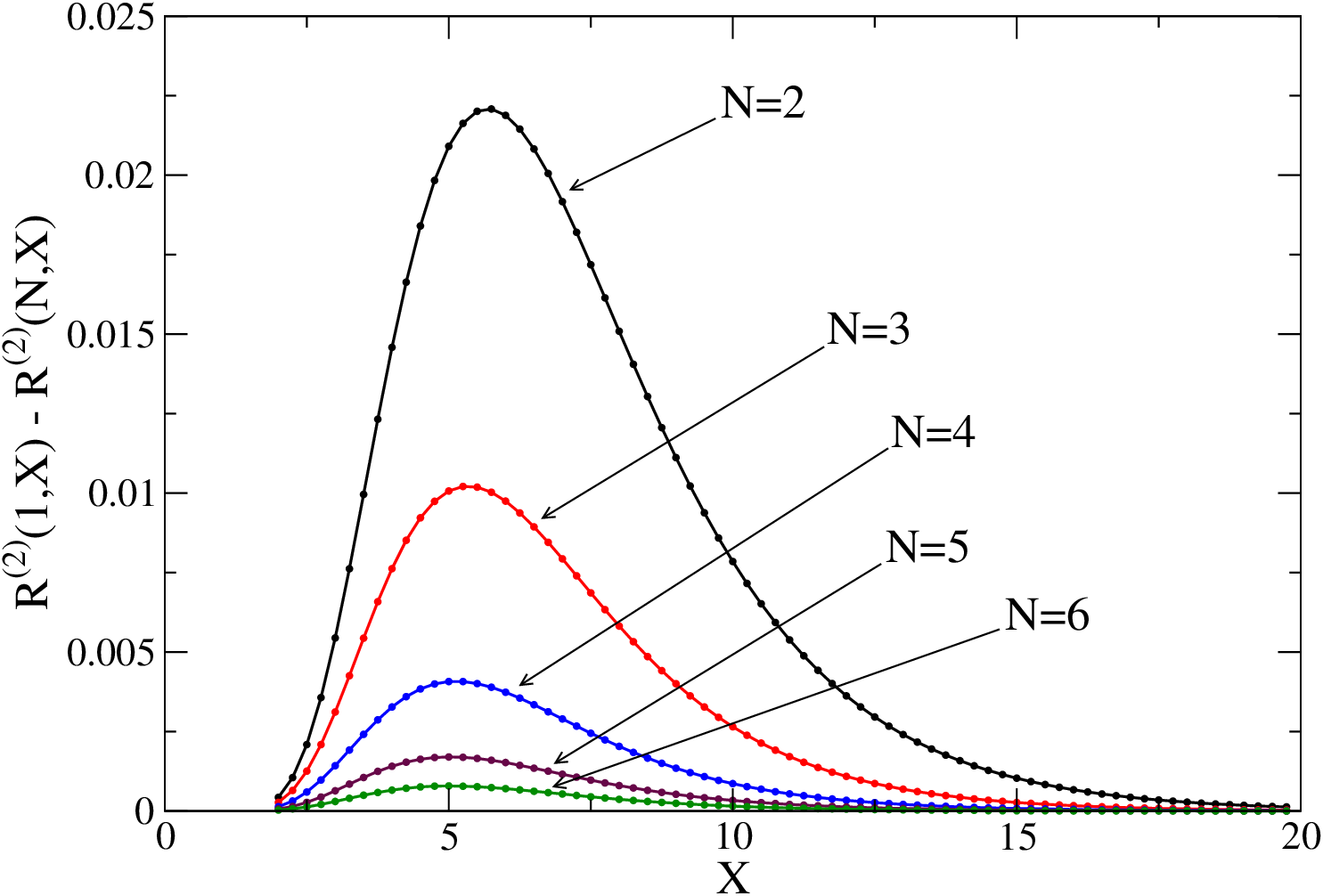}
\caption{Large $N$ behaviour of the topological susceptibility for $g=2$:
deviations of $R^{(2)}(N,X)$ from $R^{(2)}(1,X)$ are shown for $N$ up to $6$.}
\label{fig:g2}
\end{figure}

As a matter of fact, the convergence to $R(X)$ is exponentially fast in the
parameter $g$, and for $g>1$ the deviation from the above described asymptotic
value is almost irrelevant even for very small values of $N$, see
Fig.~\ref{fig:g2} for the case of $g=2$.  The really interesting cases are
therefore the spherical and toroidal topologies of the manifold, and especially
the case $g=0$, in which case (for $\theta=0$) the system is known to undergo a
finite volume phase transition transition in the large $N$ limit
\cite{Douglas:1993iia}.

\section{The large $N$ limit}\label{sec:largeN}

In this section we want to investigate the large $N$ behaviour of the
topological susceptibility and, as previously anticipated, the most interesting
case will be the $g=0$ case, since in \cite{Douglas:1993iia} a third order phase
transition was shown to be present (for $\theta=0$) in the large $N$ limit of
continuum $U(N)$ gauge theories for $g=0$. This Douglas-Kazakov transition
separates a ``small area'' region from a ``large area'' one, and it is located
at $X=\pi^2$. 

In trying to extend the Douglas-Kazakov approach to the $\theta\neq 0$ case,
one could think of writing a large $N$ effective action starting from the
partition function in Eq.~\eqref{eq:continuumpart} and using ${\hat
\theta}=\theta/N$ as scaling variable (as was done e.g. in
\cite{Bonati:2019ylr, Rossi:2016uce} following the original proposal of
\cite{Witten:1980sp}). This approach seems however problematic: the
contributions of representations corresponding to $\{l_i\}$ and $\{l_i+n\}$
(i.e. differing just for a $U(1)$ factor) differ in the large $N$ limit just
by sub-leading terms, but in the thermodynamic limit the topological
susceptibility coincides with that of the $U(1)$ model, and we can not expect
the $U(1)$ degrees of freedom to be irrelevant. It thus seems more appropriate
to construct an effective action for $\tilde W^{(g)} (N,X,k)$, and then use
Eq.~\eqref{eq:zetaweak}.

Introducing the continuous variable $y = i/N$ running from $-1/2$ to $1/2$, and
the (decreasing) function $m(y) = m_i/N$, we may define the distribution $\rho(m) =
-\mathrm{d}y/\mathrm{d}m$ and the large $N$ functional $S_{\mathrm{eff}}^{(g)}$
given by
\begin{equation}\label{eq:Seff} 
\begin{aligned} 
& S_{\mathrm{eff}}^{(g)} [m;X,k] \equiv
-\lim_{N\to\infty}\frac{1}{N^2} \ln \tilde W^{(g)} (N,X,k) =\\ 
& =(g-1)\left(\int \rho(m)\rho(m') \ln|m-m'|\mathrm{d}m\,\mathrm{d}m' +\frac{3}{2}\right)+\\ 
& +\frac{1}{2} X\left( \int \rho(m) m^2\mathrm{d}m -\bar{m}^2-\frac{1}{12}\right)
- 2 \pi i \hat{k}\, \bar{m}  \ , 
\end{aligned} 
\end{equation} 
where we defined 
\begin{equation} 
\hat{k}\equiv \frac{k}{N}\ ,\quad \bar{m} \equiv \int m\,\rho(m)\mathrm{d}m 
\end{equation} 
in order to simplify the notation. In \cite{Douglas:1993iia} the integration
domain had to be dynamically defined by the conditions $\int \rho(m)\mathrm{d}m
= 1$ and $0\le \rho(m) \leq 1$, but now $\rho(m)$ is in
general complex.

When $g>1$ the problem is singular, since for $\rho(m) \to 1$ the value of
$S_{\mathrm{eff}}^{(g)}$ approaches $-\infty$. As a consequence, since
$\rho(m)=1$ corresponds to $m(y)=-y$ (the additive constant being fixed by
the constraint $m(1/2)=-1/2$) and thus to the trivial representation of
$SU(N)$, for $g>1$ we recover the previously described trivial limit $\tilde
W^{(g)}\to 1$ and the decoupling between $SU(N)$ and $U(1)$, a conclusion that
is fully supported by the numerical results shown in Fig.~\ref{fig:g2}.

\begin{figure}[t] 
\centering 
\includegraphics[width=0.9\columnwidth, clip]{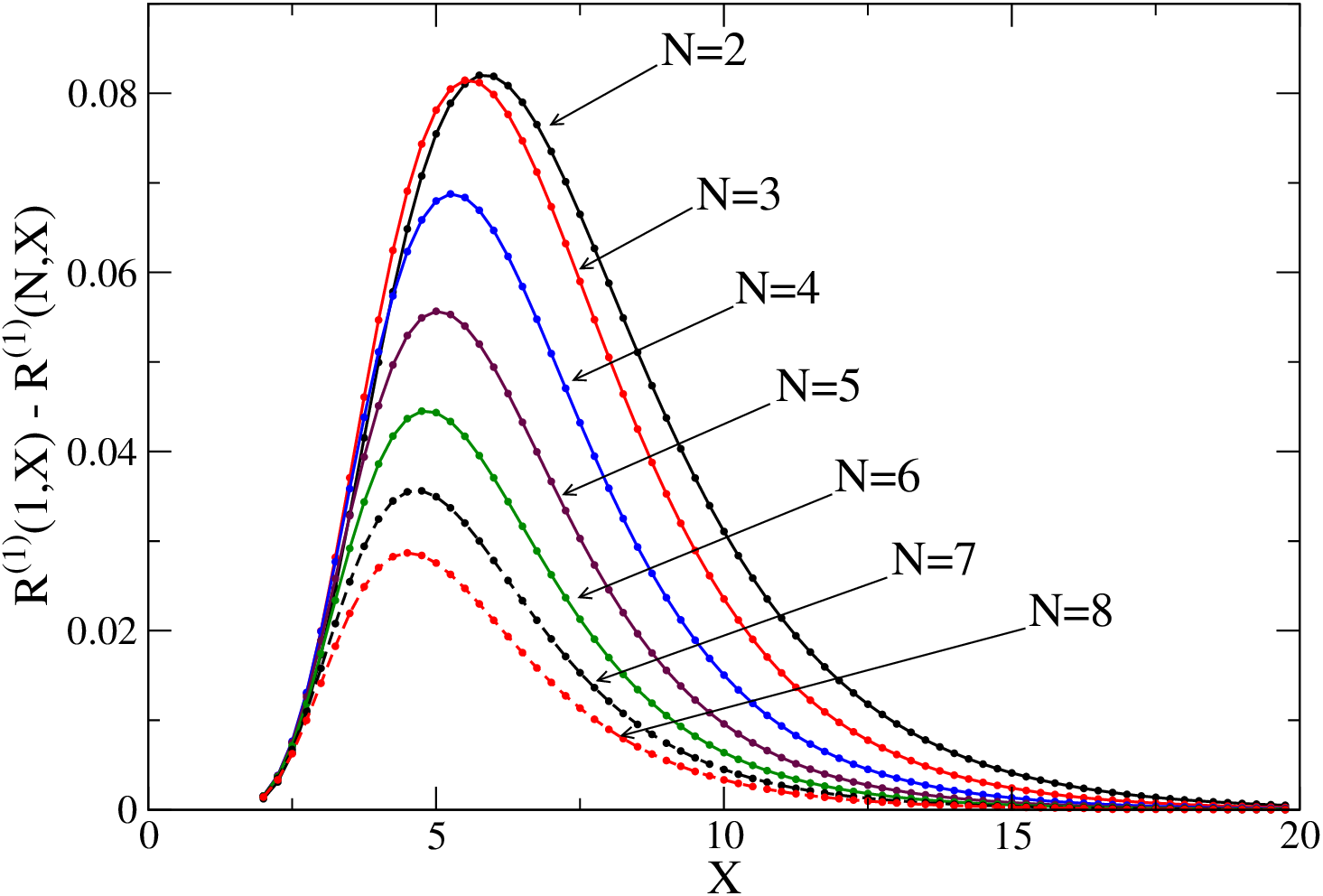}
\caption{Large $N$ behaviour of the topological susceptibility for $g=1$:
deviations of $R^{(1)}(N,X)$ from $R^{(1)}(1,X)$ are shown for $N$ up to $8$.}
\label{fig:g1}
\end{figure}

In the case $g=1$ it is known that the large $N$ expansion of the free energy
starts at order $N^0$ for $\theta=0$ (see \cite{Gross:1992tu, Gross:1993hu,
Gross:1993yt} and \cite{Douglas:1993iia}), so the basic assumption used to
obtain Eq.~\eqref{eq:Seff} is not true in this case and such an approach can
not be pursued further. One could guess, by continuity in $g$, that also in this case
the topological susceptibility in the large $N$ limit coincides with that of
the $U(1)$ case. This is strongly supported by the numerical computations
presented in Fig.~\ref{fig:g1}, where the difference
$R^{(1)}(N,X)-R^{(1)}(1,X)$ is shown (where $R^{(g)}$ is the normalized
topological susceptibility defined in Eq.~\eqref{eq:normsusc}). It is likely
that this result could be obtained directly, starting from
Eq.~\eqref{eq:continuumpart} and using the method developed in
\cite{Gross:1992tu, Gross:1993hu, Gross:1993yt}, in which case the $O(N^{-2})$
corrections of the topological susceptibiliy could maybe also be determined.

In the following we will concentrate on the analysis of the $g=0$ case, in
which case stationary points of $S_{\mathrm{eff}}^{(0)}$ are solutions of the saddle
point equation
\begin{equation}\label{eq:saddle}
-\,\mathrm{P}\!\!\int \frac{\rho(s)}{m-s}\mathrm{d}s + \frac{1}{2} X (m - \bar{m}) - i \pi\hat{k} =0 \ . 
\end{equation} 
Since we are interested just to the first $\mathcal{O}(\hat{k}^2)$
correction to the free energy, following the same approach used in
\cite{Bonati:2019ylr} we now introduce the \textit{Ansatz}
\begin{equation} 
\rho(m) = \rho_0 (m) +  i \hat{k}\, \rho_1(m)\ ,
\end{equation}
where $\rho_0(m)$ is a real even function of $m$ and $\rho_1 (m)$ is a real odd
function of $m$. The conditions $\int \rho_0(m)\mathrm{d}m = 1$ and $0\le
\rho_0(m) \leq 1$ now determine the integration domain of $\rho_0(m)$ and,
since we are interested to the leading order in $\hat{k}$, we can assume
$\rho_1(m)$ to have the same support of $\rho_0(m)$. The saddle-point equation
Eq.~\eqref{eq:saddle} thus gives for $\rho_0$ and $\rho_1$ the equations
\begin{align}
& \mathrm{P}\!\! \int \frac{\rho_0 (s)}{m-s}\mathrm{d}s = \frac{1}{2} X\,m\ , \label{eq:saddle0} \\ 
& \mathrm{P}\!\! \int \frac{\rho_1 (s)}{m-s}\mathrm{d}s = -\frac{1}{2} X \int s\,\rho_1(s) \mathrm{d}s - \pi \ .  \label{eq:saddle1} 
\end{align} 

For $X\le \pi^2$ the solution of Eq.~\eqref{eq:saddle0} is the Wigner semicircle law
\begin{equation} 
\rho_0 (m) = \frac{X}{2 \pi}\sqrt{m_0^2 -m^2}\ , \quad m_0 = \frac{2}{\sqrt{X}}\ ,
\end{equation} 
which fixes the integration domain to be $[-m_0,m_0]$.  For $X>\pi^2$ the
semicircle law would predict $\rho_0(0)>1$ and the saddle point equation
Eq.~\eqref{eq:saddle0} has to be modified, in order to make it compatible
with an \textit{Ansatz} of the form 
\begin{equation}
\rho_0(m)=\left\{\begin{array}{ll} 1                  & \mathrm{for\ } |m|\le b \\
                                   \tilde{\rho}_0(m)  & \mathrm{for\ } |m| > b  \end{array}\right.\ ,
\end{equation}
where $b$ has to be determined self-consistently, see \cite{Douglas:1993iia} for a complete discussion. 

When $X\le \pi^2$ the domain of integration to be used in
Eq.~\eqref{eq:saddle1} is thus $[-m_0,m_0]$ and this equation can be
conveniently rewritten in the form
\begin{equation}\label{eq:saddle1_bis}
\mathrm{P}\!\! \int_{-m_0}^{m_0} \frac{\rho_1 (s)}{m-s}\mathrm{d}s = C\ , \quad C=-\frac{1}{2} X \int_{-m_0}^{m_0} s\,\rho_1(s) \mathrm{d}s - \pi \ .
\end{equation} 
If we introduce as usual \cite{Brezin:1977sv} the resolvent $F(z)=\int\frac{\rho_1
(s)}{z-s}\mathrm{d}s$ 
it is simple to show that the resolvent corresponding to 
the first equation is\footnote{$\rho_1(s)$ is an odd function, so
$F(z)$ has to vanish as $z^{-2}$ for large values of $|z|$.}
\begin{equation}
F(z)=C\left (1-\frac{z}{\sqrt{z^2-m_0^2}}\right)
\end{equation}
from which it follows that 
\begin{equation}
\rho_1(m)=-\frac{C}{\pi}\frac{m}{\sqrt{m^2-m_0^2}}\ .
\end{equation}
We can now substitute this expression in the second equation in
Eq.~\eqref{eq:saddle1_bis} to fix $C$, however it is simple to show that (since
$m_0^2 X/4=1$) the resulting equation has no solution. We thus conclude that for
$g=0$ and $X\le \pi^2$ the saddle point equation Eq.~\eqref{eq:saddle1} has no
solution, and we take this fact as an indication that the topological
susceptibility vanishes in the large $N$ limit (since a nontrivial solution for
$\rho_1$ would give a susceptibility of order $N^0$).

\begin{figure}[t] 
\centering 
\includegraphics[width=0.9\columnwidth, clip]{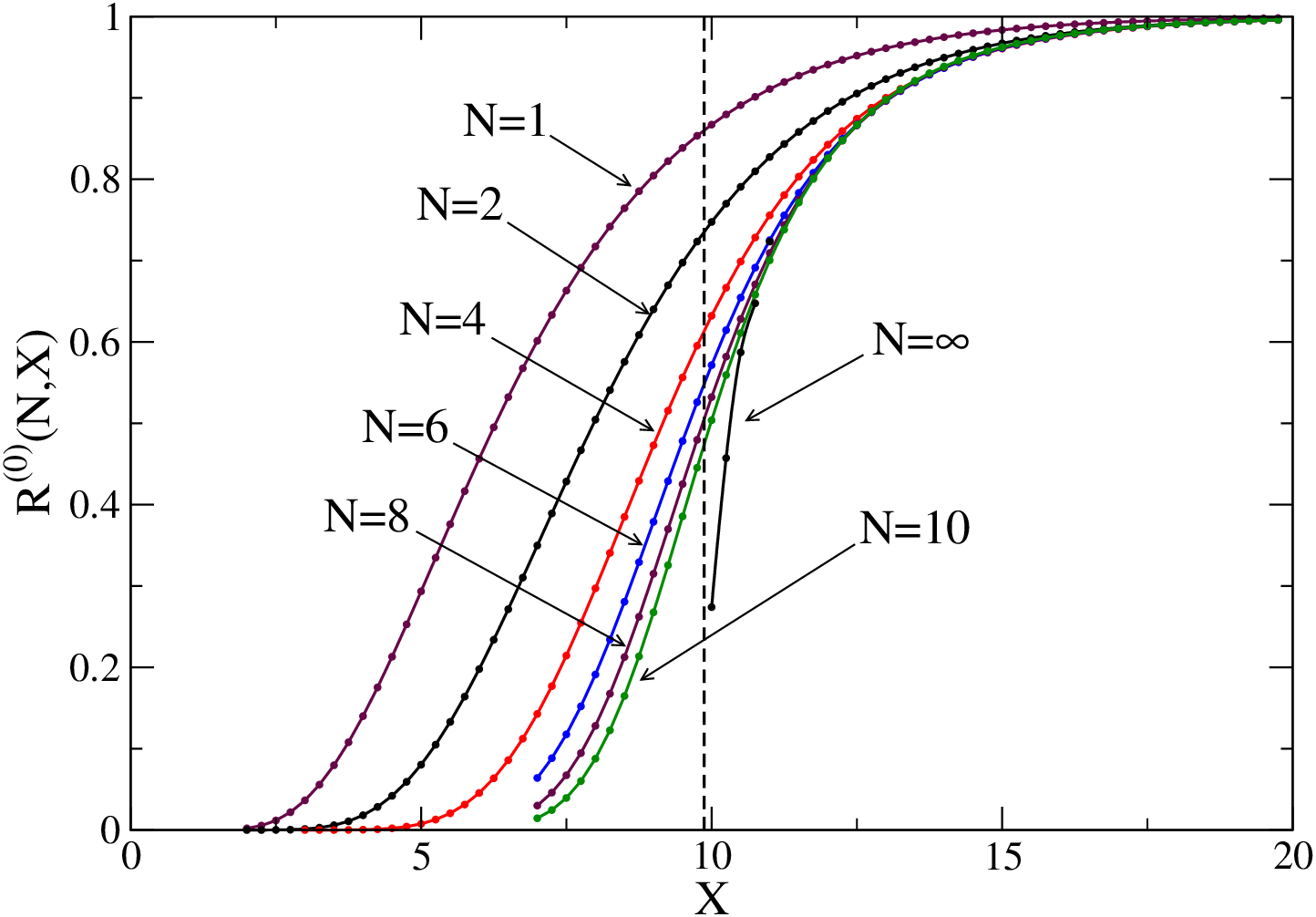}
\caption{Large $N$ behaviour of the topological susceptibility for $g=0$. The
vertical line at $X=\pi^2$ denotes the position of the Douglas-Kazakov
transition. For $10\le X \le 11$ the $N\to\infty$ extrapolation of
$R^{(0)}(N,X)$ is also shown, which is obtained from the results of Monte-Carlo
simulations performed at $N\ge 30$, see the text for more details.}
\label{fig:g0w}
\end{figure}

\begin{figure}[b] 
\centering 
\includegraphics[width=0.9\columnwidth, clip]{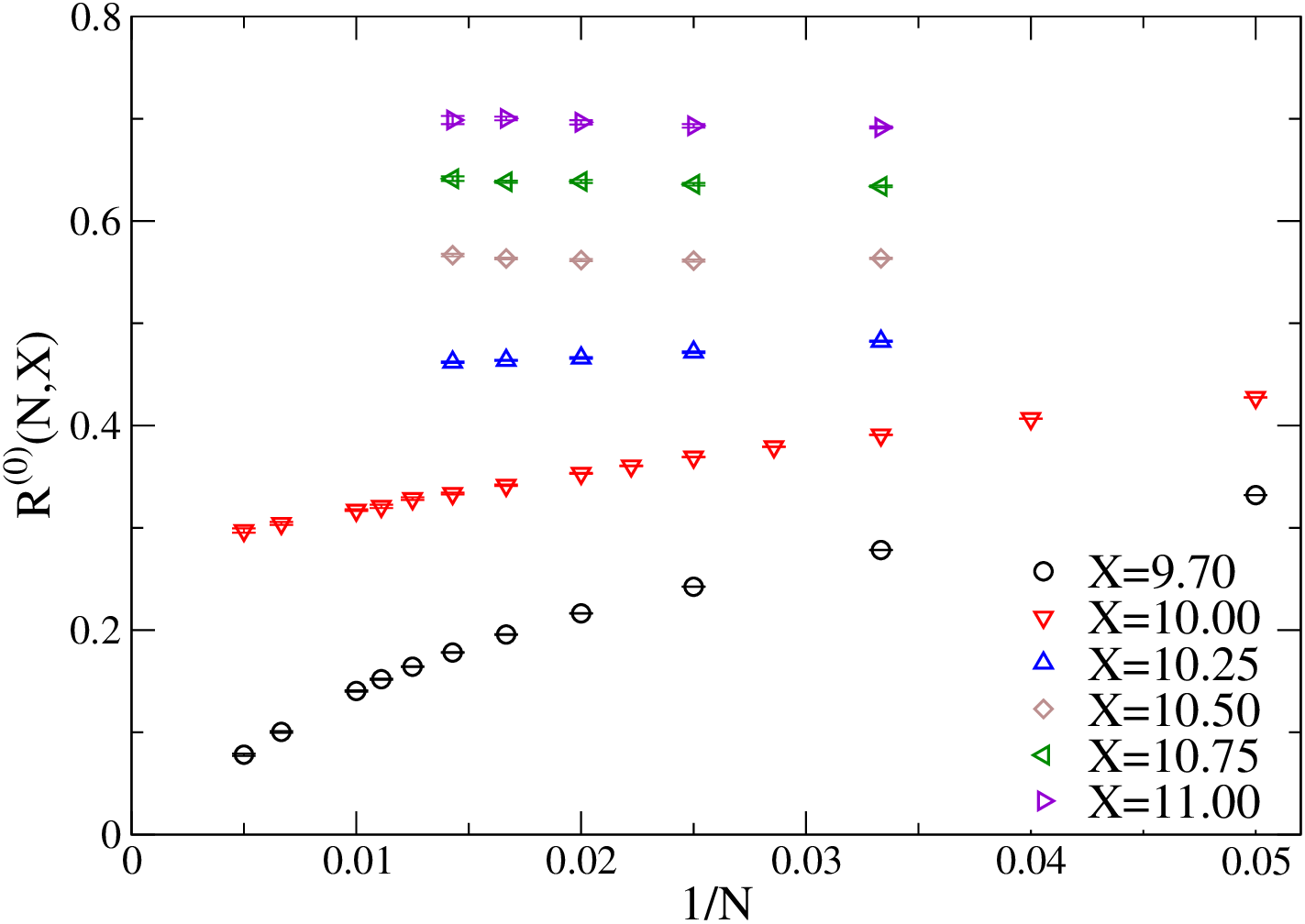}
\caption{Large $N$ behaviour of the topological susceptibility for $g=0$ and
several $X$ values. }
\label{fig:g0mc}
\end{figure}

\begin{figure}[t] 
\centering 
\includegraphics[width=0.9\columnwidth, clip]{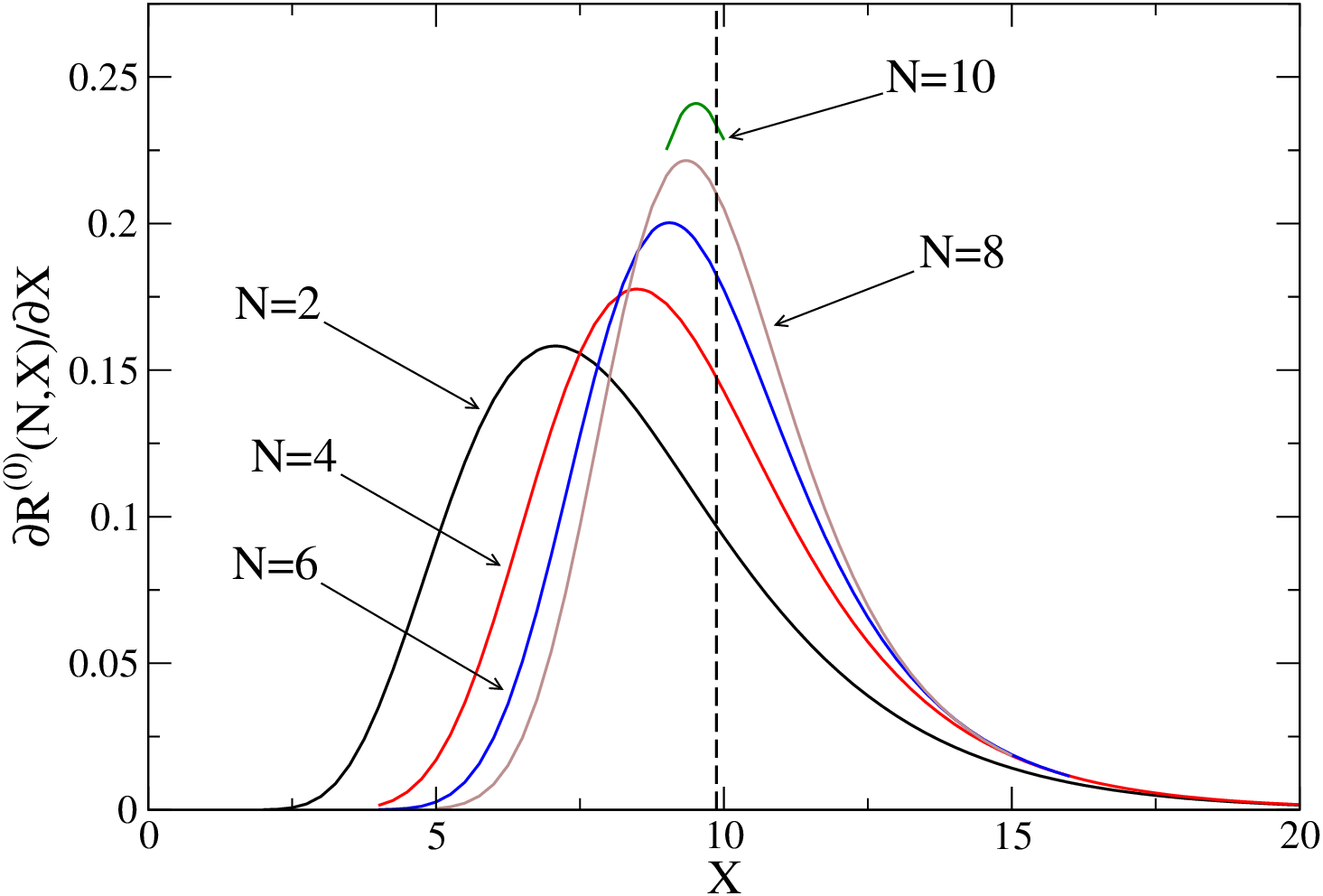}
\caption{Large $N$ behaviour of $\partial R^{(0)}(N,X)/\partial X$. For the $N$
values shown in this figure the peak seems to approach the Douglas-Kazakov
transition, indicated by the dashed vertical line.}
\label{fig:dRdx_w}
\end{figure}

For $X>\pi^2$ the saddle point equation for $\rho_0$ has to be modified in
order for its solution to satisfy the requirement $\rho_0(m)\le 1$
\cite{Douglas:1993iia}, but it is
not clear if the equation for $\rho_1$ has also to be modified (and eventually
how). In absence of a clear theoretical understanding of this point, the
following analysis will be based exclusively on numerical evidence.

In Fig.~\ref{fig:g0w} we report the behaviour of the normalized topological
susceptibility $R^{(0)}(N,X)$ (defined in Eq.~\eqref{eq:normsusc}) for
some values of $N$, up to $N=10$. It is clear that lines corresponding to
increasing $N$ values are not converging to the $N=1$ curve. For $X$ smaller
than $\pi^2$ the values of $R^{(0)}(N,X)$ seem to approach zero
as $N$ grows, while for $X$ larger than $\pi^2$
they seem to converge to nonzero values in the same limit. Around $\pi^2$ a transition region is
present, in which the behaviour of $R^{(0)}(N,X)$ rapidly changes.

These results have been obtained by explicitly performing the sums over
$\{m_i\}$ up to a prescribed relative accuracy of $10^{-6}$
(the sum on $n$ can be rewritten in term of Jacobi $\theta$-functions),
however, in order to reach larger values of $N$, we found computationally much
more efficient to estimate average values using a Monte-Carlo sampling of the
distribution in Eq.~\eqref{eq:cont_weights}. Using this approach we
obtained the data shown in Fig.~\ref{fig:g0mc}, where the large $N$ behaviour
of $R^{(0)}(N,X)$ is scrutinized for two values of $X$ close to $\pi^2\approx
9.8696$ ($X=9.7$ and $X=10$) using values of $N$ up to 200, and for larger $X$
values using $30\le N\le 70$. The large $N$ behaviour of the topological
susceptibility is consistent with the one guessed from the results obtained using
$N\le 10$, however values of $N$ larger than $50$ are needed to clearly
appreciate this behaviour for the two $X$ values closer to $\pi^2$. From these
data we extracted the large $N$ limit of $R^{(0)}(N,X)$ shown in
Fig.~\ref{fig:g0w} for $10\le X\le 11$.

Data presented so far indicate that for $g=0$ manifolds the large $N$
topological susceptibility vanishes for $X<\pi^2$ while it is nonzero for
larger values of $X$, approaching the $U(1)$ values as $X\gg 1$. From
Figs.~\ref{fig:g0w}-\ref{fig:g0mc} we can see that the transition between the
two regimes is quite abrupt, but we have no real hints on what happens at
$X=\pi^2$. To further investigate the region $X\simeq \pi^2$ it is convenient
to study $\partial R^{(0)}(N,X)/\partial X$ (see Eq.~\eqref{eq:dRdx} for the
explicit expression of this quantity), in order to understand if this
observable develops a singularity at $X=\pi^2$ as $N$ gets larger.

\begin{figure}[t] 
\centering 
\includegraphics[width=0.9\columnwidth, clip]{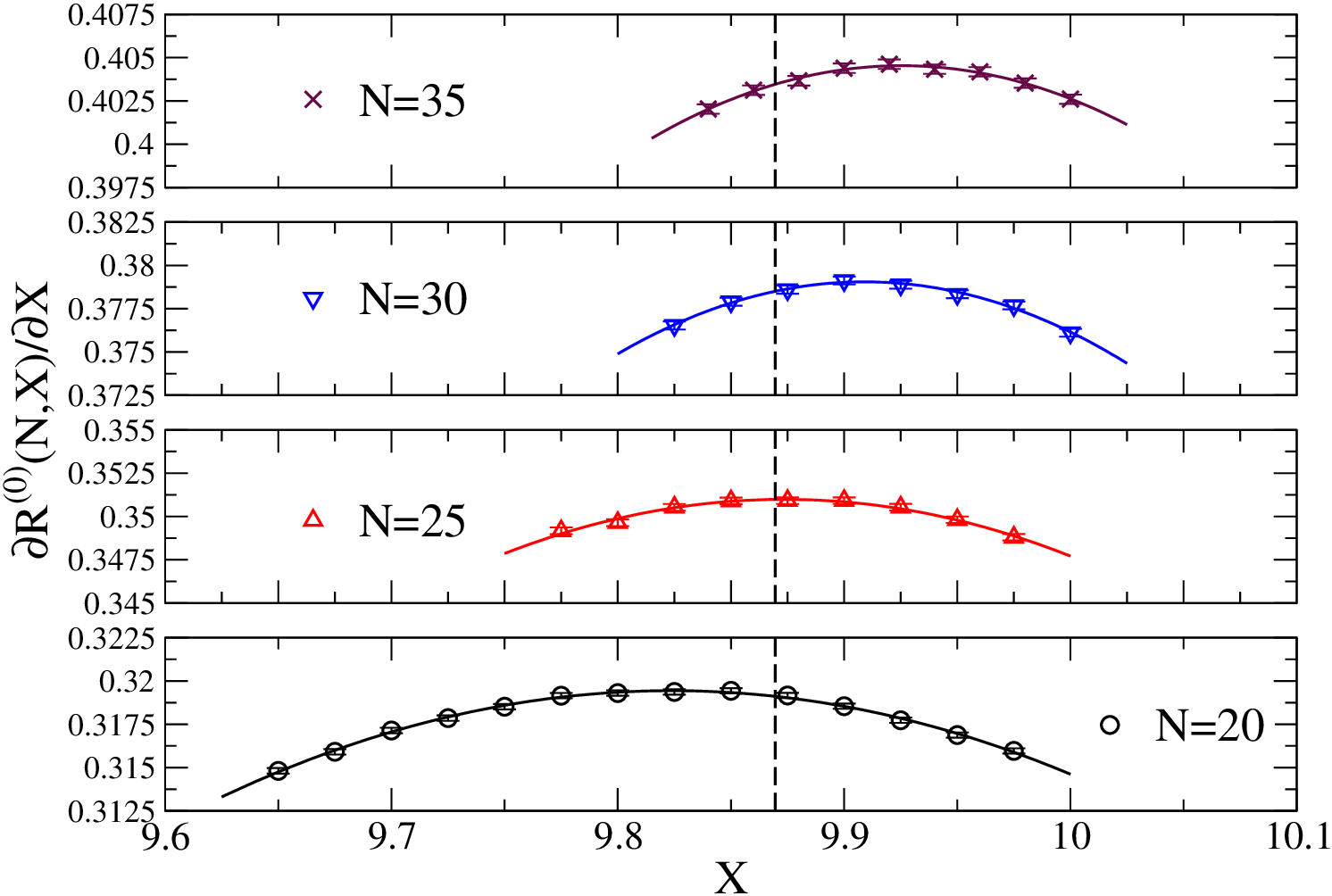}
\caption{Large $N$ behaviour of $\partial R^{(0)}(N,X)/\partial X$. The
dashed vertical line denotes the location of the Douglas-Kazakov transition and it is
clear that the peak moves into the large area phase. Continuous lines are
quadratic fit and they are drawn just to guide the eye.}
\label{fig:dRdx_mc}
\end{figure}

In Fig.~\ref{fig:dRdx_w} the profile of $\partial R^{(0)}(N,X)/\partial X$ is
shown for some $N$ values up to $N=10$, and a singularity at $X=\pi^2$ indeed
seems to emerge. In order to reach larger $N$ values and better investigate
this ``might be'' singular behaviour we again resorted to Monte-Carlo
simulations, and the results obtained in this way are shown in
Fig.~\ref{fig:dRdx_mc}.  By looking just at data corresponding to $N\lesssim
25$ one could guess that the position of the peak of $\partial
R^{(0)}(N,X)/\partial X$ approaches $\pi^2$, however data at larger $N$ show
that the peak crosses the Douglas-Kazakov transition, going into the large-area
regime. This is consistent with a continuous behaviour of $\partial
R^{(0)}(N,X)/\partial X$ at the transition at $X=\pi^2$. Note however that this
behaviour is formally continuous but nonetheless very abrupt, indeed for
$X\lesssim 35$ the peak-value of $\partial R^{(0)}(N,X)/\partial X$ is still
growing almost linearly in $N$ and its location is still very close to
that of the Douglas-Kazakov transition.

\section{Conclusions}\label{sec:conclusions}

In this paper we investigated the finite volume $\theta$-dependence of
continuum two dimensional $U(N)$ gauge theories. We previously noted in
\cite{Bonati:2019ylr} that at finite volume the $U(1)$ degrees of
freedom do not factorize in the partition function of 2d $U(N)$ gauge theories,
even in the continuum limit. The continuum partition function was however
written in a way that made the form of the interaction between the $U(1)$ and
the $SU(N)$ degrees of freedom non completely clear. 

In the present work we showed that the $\theta$-dependent continuum partition
function can be rewritten in the more transparent form in
Eq.~\eqref{eq:zetaweak}. In this new form $\theta$ couples only to the $U(1)$
instanton number, but the effective action of the $SU(N)$ degrees of freedom
generically depends on the topological charge of the background $U(1)$ field.
In some specific limits, like in the thermodynamical limit ($X\to\infty$) or in
the large genus limit ($g\to\infty$), this dependence disappears; only in these
cases the $\theta$-dependence of the continuum 2d $U(N)$ theory reduces to that
of the continuum $U(1)$ theory.

We then investigated the large $N$ behaviour of the topological susceptibility,
mainly by means of numerical simulations. We found that, in the large $N$ limit
and for fixed area of the manifold, the topological susceptibility converges to
its $U(1)$ value only if the genus of the manifold is larger than zero.

In the case of a manifold with the topology of the sphere, the large $N$
topological susceptibility turned out to be an order parameter for the
Douglas-Kazakov transition at $\theta=0$ \cite{Douglas:1993iia}: the large $N$
limit of the topological susceptibility vanishes in the small-area phase
$X<\pi^2$ and it is different from zero in the large-area phase.
Moreover the derivative with respect to the area of the topological
susceptibility is continuous across the transition. 

This behaviour is the analogous, in the continuum finite area case, of the one
previously found in \cite{Bonati:2019ylr}, where the large $N$ behaviour of the
topological susceptibility was shown to be different in the two phases of the
Gross-Witten-Wadia transition. However for the case studied in
\cite{Bonati:2019ylr} an explicit analytic expression for the large $N$
topological susceptibility was found, while in the present case we had to rely
mostly on numerics.

\acknowledgments

Numerical computations have been performed by using resources provided by the
Scientific Computing Center at INFN-PISA.

\end{document}